# Comparative first-principles studies of prototypical ferroelectric materials by LDA, GGA, and SCAN meta-GGA

Yubo Zhang,[1,2] Jianwei Sun,[2,*] John P. Perdew,[1,3] and Xifan Wu[1,*]

[1]*Department of Physics, Temple University, Philadelphia, Pennsylvania 19122, USA*
[2]*Department of Physics, University of Texas at El Paso, El Paso, Texas 79968, USA*
[3]*Department of Chemistry, Temple University, Philadelphia, Pennsylvania 19122, USA*

Originating from a broken spatial inversion symmetry, ferroelectricity is a functionality of materials with an electric dipole that can be switched by external electric fields. Spontaneous polarization is a crucial ferroelectric property, and its amplitude is determined by the strength of polar structural distortions. Density functional theory (DFT) is one of the most widely used theoretical methods to study ferroelectric properties, yet it is limited by the levels of approximations in electron exchange-correlation. On the one hand, the local density approximation (LDA) is considered to be more accurate for the conventional perovskite ferroelectrics such as $BaTiO_3$ and $PbTiO_3$ than the generalized gradient approximation (GGA), which suffers from the so-called *super-tetragonality* error. On the other hand, GGA is more suitable for hydrogen-bonded ferroelectrics than LDA, which largely overestimates the strength of hydrogen bonding in general. We show here that the recently developed general-purpose strongly constrained and appropriately normed (SCAN) meta-GGA functional significantly improves over the traditional LDA/GGA for structural, electric, and energetic properties of diversely-bonded ferroelectric materials with a comparable computational effort, and thus enhances largely the predictive power of DFT in studies of ferroelectric materials. We also address the observed system-dependent performances of LDA and GGA for ferroelectrics from a chemical bonding point of view.

## 1. INTRODUCTION

Ferroelectricity is an important property of materials (e.g., ferroelectric materials $BaTiO_3$ and $PbTiO_3$, and multiferroic material $BiFeO_3$) that have a spontaneous electric polarization below the Curie temperature, and the polarization direction is switchable when the applied electric field is greater than the coercive field.[1,2] Microscopically, the polarization is induced by the breaking of spatial inversion symmetry of the crystal. Understanding the driving force or microscopic mechanism of the symmetry breaking lies at the heart of the development of novel high-performance candidates. From the one-electron perspective the mechanism of ferroelectric polarization is orbital hybridization,[3] while from the many-electron perspective the mechanism is competition between the exchange-correlation (XC) energy, which favors more inhomogeneous densities, and the rest of the total energy. For example, symmetry breaking in the $BaTiO_3$ ferroelectric phase is driven by a zone-center lattice instability of the centrosymmetric (paraelectric) phase, which, at the atomic level, demonstrates collective displacements of Ti ions away from $TiO_6$ octahedra centers. The amplitude of the displacement is thereby an important factor in determining the ferroelectric properties. Alternatively, the polar distortion in ferroelectrics can also have an electronic origin (e.g., $LuFe_2O_4$).[4]

Density functional theory (DFT) is a powerful theoretical tool for studying ferroelectric properties. DFT can be used to calculate not only structural and dynamical[5,6] properties for ferroelectrics but also the Berry phase polarization according to the modern theory of polarization.[7] Although the DFT approaches have been routinely used for studying ferroelectric materials, the predicted ferroelectric properties sensitively depend on the adopted exchange-correlation functionals,[8,9] such as the local density approximation (LDA),[10] the semilocal generalized gradient approximation (GGA) in the standard form of the Perdew–Burke–Ernzerhof (PBE),[11,12] and the non-local hybrid functionals (e.g., PBE0,[13-16] B3LYP,[17-20] and HSE[21,22]) that mix semilocal with exact exchange. For the most-studied perovskites $BaTiO_3$ and $PbTiO_3$, for example, the LDA-predicted ferroelectric lattice distortion (i.e., $c/a > 1$), spontaneous polarization, and lattice dynamics agree well with experimental results.[8,9] By contrast, PBE is less often used because of the strong overestimation of lattice distortion, which is known as the *super-tetragonality* problem.[9] Interestingly, PBE is more reliable than LDA for the studies of hydrogen-bonded systems,[23] for which LDA severely overestimates the strength of the hydrogen bonds.[24,25] The hybrid functionals usually can give improved structural properties at the expense of considerably increased computational cost. The widely used Heyd-Scuseria-Ernzerhof (HSE) functional,[21,22] however, still significantly overestimates the structural distortion for the perovskites, inheriting the *super-tetragonality* problem from its parent PBE functional.[26] To avoid the above errors, a hybrid functional named B1-WC that hybridizes 16% of the exact exchange was specifically designed for ferroelectric materials.[9,27,28] A substantial improvement was observed for the B1-WC hybrid functional but is restricted to a few small systems so far.

---

[*]Authors to whom correspondence should be addressed. Electronic addresses: xifanwu@temple.edu and jsun2@utep.edu

The system-dependent performances of the above-mentioned XC functionals, due to lack of a universal treatment for diversely-bonded ferroelectrics, strongly limit the predictive power of DFT especially for exploring new materials. The recently developed strongly constrained and appropriately normed (SCAN) meta-GGA,[26,29] has been shown to systematically improve over LDA/PBE for geometries and energies of diversely-bonded materials (including covalent, metallic, ionic, hydrogen, and van der Waals bonds), and thus to enhance largely the predictive power of DFT. In Ref. 26 SCAN has been shown to systematically improve over LDA/PBE, and is often as or more accurate than the hybrid B1-WC for the structural properties and spontaneous polarizations of $BaTiO_3$, $PbTiO_3$, and $BiFeO_3$. SCAN has also been shown to improve the ferroelectric transition temperatures in $BaTiO_3$, $KNbO_3$, and $PbTiO_3$.[30] The computational cost of a meta-GGA such as SCAN is moderately greater than that of LDA or GGA, but significantly less than that of a hybrid functional.

In this paper, we perform a comparative investigation of various prototypical ferroelectric materials. Our selected systems can be roughly classified into (1) perovskite systems ($BaTiO_3$, $PbTiO_3$, and $LiNbO_3$), (2) hydrogen-bonded systems (inorganic $KH_2PO_4$ and organic 2-phenylmalondialdehyde), and (3) multiferroic systems ($BiFeO_3$ and $YMnO_3$). We first perform a brief survey on the bonding interactions that are involved in each system. In the following sections, the structural and ferroelectric properties are comparatively investigated using LDA, PBE, and SCAN, as well as the hybrid functionals HSE and B1-WC. Our results show that the SCAN functional is a universally accurate approach for the selected systems. This systematic improvement in performance is attributed to the systematic construction of SCAN to satisfy all known exact constraints that its flexible functional form can satisfy, and beyond that to fit appropriate norms: non-bonded systems for which its form can be expected to be highly accurate.

## 2. COMPUTATIONAL DETAILS

The structural properties and ferroelectric polarization are calculated using the Vienna Ab-initio Simulation Package (VASP)[31] with the projector-augmented wave method.[32,33] The LDA,[34,35] the GGA in the form of the PBE,[11,12] and the SCAN[26,29] meta-GGA[36] are used for comparative studies. The hybrid functional HSE[21,22] is also applied to selected systems for comparison. The semicore *p*-states are taken as valence states for Ti, Nb, Mn, and Fe; semicore *d*-states are taken as valence states for Pb and Bi. Taking the semicore *s*-states (e.g., Ti-3*s*, Nb-4*s*, and O-2*s*) as valence electrons only has a slight influence, and these states are treated as core states for simplicity. An energy cutoff of 600 eV is used to truncate the plane wave basis. We use Γ-centered 8 × 8 × 8 *K*-meshes for the five-atom-cell of $BaTiO_3$ and $PbTiO_3$, and 4 × 4 × 2 *K*-meshes for the 30-atom-cell of $BiFeO_3$ and $YMnO_3$. The spin configuration is G-type antiferromagnetic for $BiFeO_3$[37] and A-type for $YMnO_3$.[38] The spin-orbit coupling effect is neglected for all the systems. Crystal structures are fully relaxed (with a force convergence criterion of 0.001 eV/Å) unless otherwise stated. The spontaneous polarization is calculated according to the modern theory of polarization.[7] The vibrational frequencies at the Brillouin zone center are computed using the density perturbation functional theory.[5,6] The phonon dispersion relations (for $BaTiO_3$, $PbTiO_3$, and $SrTiO_3$) in the full Brillouin zone are calculated using the frozen phonon approach with 3 × 3 × 3 supercells using the Phonopy package.[39]

## 3. RESULTS AND DISCUSSIONS

### 3.1. A survey of diverse bonding interactions in prototypical ferroelectric materials

The conventional $BaTiO_3$, $PbTiO_3$, and $LiNbO_3$ oxides, the hydrogen-bonded $KH_2PO_4$ and organic 2-phenylmalondialdehyde (PhMDA), and the multiferroic $BiFeO_3$ and $YMnO_3$ oxides are chosen as examples (see Figure 1 for the crystal structures). For all these materials, the ferroelectric structural distortions are induced by, if viewed from the one-electron perspective, electronic hybridization[3] between the transition metal and oxygen atoms in the oxides, or between hydrogen and oxygen ions in the hydrogen-bonded systems. Considering the diversity of the selected materials, it is necessary to briefly introduce the bonding interactions involved in each type of systems.

The ferroelectric properties of the perovskite $BaTiO_3$, $PbTiO_3$, and $LiNbO_3$ have been extensively studied in the past. In $BaTiO_3$, the polar distortions can be described by a collective off-centering displacement of Ti ions,[3] which results in shorter Ti-O bonds along one direction and longer Ti-O bonds along the opposite direction [see Figure 1(a)]. While the Ba-O bond is essentially ionic,[3] the Ti-O ionic-like interaction also mixes considerable covalent bonding character.[3,40] The short-range electrostatic repulsion between the electron clouds on the adjacent Titanium and Oxygen ions, which favors the paraelectric structure, can be softened by the covalent hybridization between the O-2*p* and the nominally empty Ti-3$d^0$ orbitals (i.e., the $d^0$-ness requirement), which tends to stabilize the tetragonal ferroelectric phase.[41] As a consequence, the O-2*p* Ti-3$d^0$ covalent hybridization is closely associated with the ferroelectric structural instability.[3,42] From a chemical bonding point of view, the amplitude of the ferroelectric distortion can be influenced by the fraction of involved covalent and ionic characteristics. It is worth noting that the actual amplitude depends on terms in the free energy that vary with different degrees of freedom, e.g., anharmonicity and strains. In $PbTiO_3$, in addition to the Ti distortion, the Pb off-center displacement also contributes significantly to the ferroelectric distortion due to the covalent hybridization between the stereochemically active Pb-6$s^2$ lone-pair electrons and O-2*p* electrons.[43] Therefore, both the *pd* (i.e., O-2*p* and Ti-3*d*) and *sp* (i.e., Pb-6*s* and O-2*p*) hybridizations should be reliably captured to describe the ferroelectric properties accurately. For the perovskite-like $LiNbO_3$, the ferroelectric phase transition involves two kinds of structural distortions.[44,45] Both the ordering of $Li^{1+}$ ions to a particular side of the oxygen planes along the *c*-axis and the displacement of $Nb^{5+}$ ions away from the $NbO_6$ centers towards the upper facets [see Figure 1(b)] contribute to the ferroelectric polarization.[44,45] Nevertheless, the amplitude of the ferroelectric distortion is strongly influenced by the $Nb^{5+}$ off-center displacement, which is associated with the hybridization between the Nb-4$d^0$ and O-2*p* orbitals.[44]

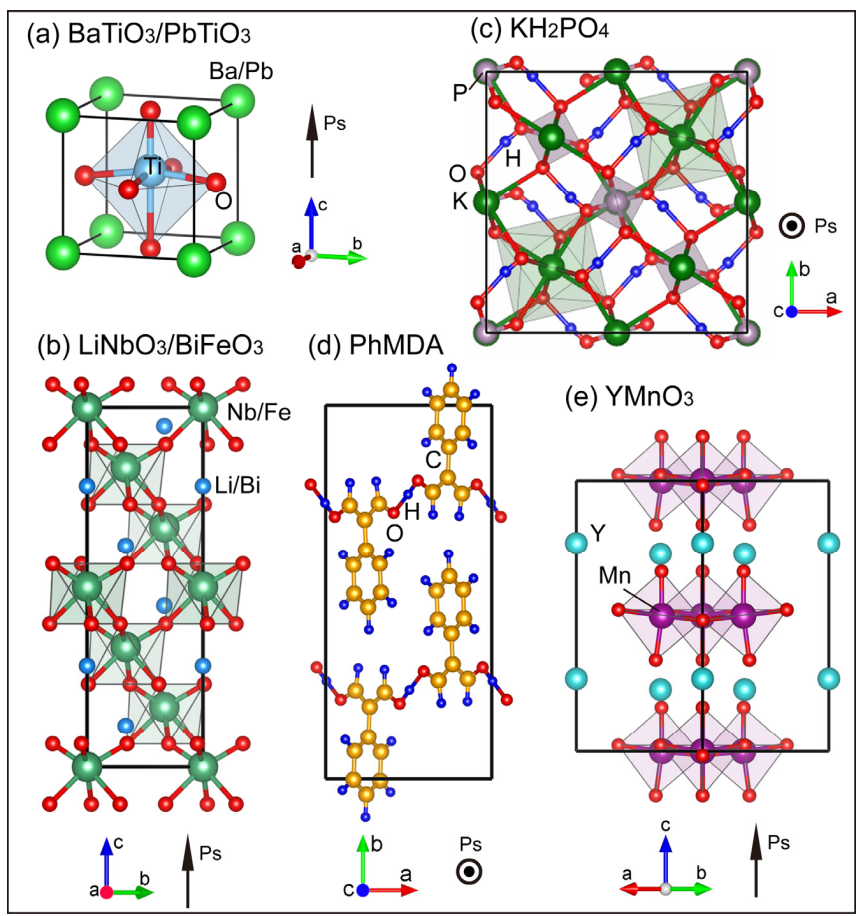

Figure 1. The low-temperature ferroelectric phase of several prototypical ferroelectric materials. (a) Tetragonal BaTiO$_3$ and PbTiO$_3$ with space group *P4mm*, (b) perovskite-like LiNbO$_3$ and BiFeO$_3$ with space group *R*3*c*, (c) orthorhombic KH$_2$PO$_4$ with space group *Fdd*2, (d) orthorhombic 2-phenylmalondialdehyde (C$_9$H$_8$O$_2$, known as PhMDA) with space group *Pna*2$_1$, and (e) hexagonal YMnO$_3$ with space group *R*3*c*. The black arrows show the spontaneous polarization directions with respect to lattices. Note that the O−H⋯O bonds in KH$_2$PO$_4$ are almost within the *ab*-plane (i.e., the basal plane), but the bonds in PhMDA have both the in-plane and out-of-plane components.

Multiferroic materials have great potential in lots of applications since they combine the electric and magnetic degrees of freedom in a single material. The perovskite BiFeO$_3$ and hexagonal YMnO$_3$ are the two well-studied systems. In BiFeO$_3$, because the partial occupation of Fe ions breaks the $d^0$-ness rule, the ferroelectric distortion is no longer associated with the *pd* hybridization (as between O-2*p* and Ti-3*d* states in BaTiO$_3$). Instead, the ferroelectric distortion is dominated by the Bi collective displacement towards the oxygen atoms along lattice direction *c* [see Figure 1(b)], which is related to the hybridization between the Bi-6$s^2$ lone-pair electrons and the O-2*p* electrons.[43] The induced off-center displacements of Fe atoms in the FeO$_6$ octahedra also contribute to the polarization but play a minor role compared to the former.[46] However, the partially occupied Fe-3*d* orbitals are directly related to the magnetic properties. For example, the Néel temperature and magnetic moment can be sensitively influenced by the Fe-O interaction. Therefore, a reliable prediction of the ferroelectric, magnetic, and magnetoelectric coupling properties requires accurate descriptions of the *sp* (i.e., Bi-6*s* and O-2*p*) and the *pd* (i.e., O-2*p* and Fe-3*d*) hybridizations. The magnetoelectric coupling in BiFeO$_3$ is, however, expected to be weak because the electric and magnetic orderings originate from two different sources related to Bi and Fe atoms, respectively.[47] By contrast, the coupling in hexagonal YMnO$_3$, which is known as an *improper* ferroelectric material, can be more efficient. This is due to the mutual dependence of two phonon modes (i.e., the *improper* nature). On one hand, the ferroelectric distortion (i.e., related to a $\Gamma_2^-$ mode) is indirectly induced by a non-ferroelectric mode (i.e., $K_3$) described as the rotation of MnO$_5$ triangular bipyramids.[48] On the other hand, the $K_3$ mode, which is directly coupled to the weak ferromagnetic spin canting, can be driven by the $\Gamma_2^-$ mode within an external electric field.[49] At the atomic level, the stabilization of the ferroelectric phase is associated with the O-2$p$ Y-4$d$ hybridization with the $d^0$-ness requirement being fulfilled by Y$^{3+}$ ions,[50,51] and the magnetic property is determined by the Mn-3$d$ electronic states hybridized with O-2$p$ orbitals. As a result, reliable predictions of the two types of $pd$ hybridizations, i.e., between the O-2$p$ states with both the Y-4$d$ and the Mn-3$d$ states, are essential for the calculation of the YMnO$_3$ multiferroic properties.

Whereas the inorganic KH$_2$PO$_4$ (known as KDP) is one of the earliest found ferroelectric materials with hydrogen bonds, the organic supramolecular systems[52,53] [e.g., 1-cyclobutene-1,2-dicarboxylic acid (C$_6$H$_6$O$_4$, known as CBDC)[54] and 2-phenylmalondialdehyde (C$_9$H$_8$O$_2$, known as PhMDA)[55]] are also recently theoretically explored[23] for their ferroelectric properties. In KH$_2$PO$_4$, it is well known that the polar distortion and, therefore, the spontaneous polarization are related to the proton ordering below the Curie temperature. Nevertheless, the polarization direction [along the lattice *c*, see Figure 1(c)] is perpendicular to the proton displacement directions (within the basal plane). Theoretical studies[56,57] revealed that the polarization is due to both the (Hydrogen) electronic charge redistributions and the (Phosphorus) ionic displacements, both of which critically depend on the proton ordering. Similarly, the polar distortion in the organic ferroelectrics [e.g., the PhMDA as shown in Figure 1(d)] is also closely related to the proton ordering. In both KH$_2$PO$_4$ and PhMDA, the essence of proton ordering is the coherent alignment of the two types of inequivalent bonds, i.e., the O-H covalent bonds and the O···H hydrogen bonds; the amplitude of ferroelectric distortions is determined by the competition between the O-H and the O···H bonding interactions.

The above discussions clearly indicate that diverse bonding interactions are involved in generating the ferroelectric distortions. Theoretically, it is challenging to have predictions agreeing well with experiments when there are competing interactions, and it is therefore highly desirable to have an efficient density functional that can treat these bonding interactions with similar accuracy. In the following sections, we present comparative studies on the examples mentioned above using the LDA, PBE, SCAN, and hybrid functionals (e.g., HSE and B1-WC) to show that SCAN meets all the requirements.

## 3.2. Perovskite ferroelectrics with $d^0$-ness ions: BaTiO$_3$, PbTiO$_3$, and LiNbO$_3$

Perovskite ferroelectrics such as BaTiO$_3$, PbTiO$_3$, and LiNbO$_3$ with $d^0$-ness ions are among the most extensively studied systems. We calculate the structural properties (lattice parameter *a*, lattice distortion ratio or tetragonality *c/a*, cell volume, displacement of transition metal ions with respect to the centrosymmetric structure), the ferroelectric properties (Born effective charge, spontaneous polarization, the energetic difference between the paraelectric and the ferroelectric phases), and the band gaps of BaTiO$_3$, PbTiO$_3$, and LiNbO$_3$ using various popular XC functionals. The results are shown in Table 1 and are also plotted in Figure 2. The previously computed results by B1-WC[9] and the available experimental values are also presented for comparison.

Table 1. Structural, ferroelectric, and electronic properties of $P4mm$ BaTiO$_3$, $P4mm$ PbTiO$_3$, and $R3c$ LiNbO$_3$. Structural properties include the lattice parameter $a$ (Å), tetragonality $c/a$, unit cell volume $\Omega$ (Å$^3$), and atomic displacement $\Delta_{Ti}$ or $\Delta_{Nb}$ (in units of the lattice constant $c$). Ferroelectric properties include the Born effective charge $Z^*$ of Ti/Nb and O (in directions perpendicular and parallel to the lattice direction $c$), spontaneous polarization $P_s$ (µC/cm$^2$), and energetic (eV) difference between paraelectric (PF) and ferroelectric (FE) phases. Note that hexagonal unit cell is used for LiNbO$_3$ in the calculations. The paraelectric phase is used for the $Z^*$ calculation in order to compare with a previous report.[9] The PP and AE stand for the pseudopotential and the all-electron potential, respectively. The results are also plotted in Figure 2 for easier comparison. Some results for BaTiO$_3$ and PbTiO$_3$ are reproduced from Ref. 26.

| Approaches | Lattice $a$ (Å) | Tetragonality $c/a$ | Volume $\Omega$ (Å$^3$) | Displacement $\Delta_{Ti}$ or $\Delta_{Nb}$ | Born charge $Z^*_{Ti\ or\ Nb}$, $Z^*_{O_{\perp c}}/Z^*_{O_{\parallel c}}$ | Polarization $P_s$ (µC/cm$^2$) | $E_{PF}-E_{FE}$ $\Delta E$ (meV/atom) | Band gap $E_g$ (eV) |
|---|---|---|---|---|---|---|---|---|
| | | | | BaTiO$_3$ | | | | |
| LDA | 3.946 | 1.011 | 62.1 | 0.012 | 7.44, −2.17/−5.85 | 24.3 | 1.0 | 1.72 |
| PBE | 4.000 | 1.054 | 67.5 | 0.018 | 7.52, −2.13/−5.99 | 47.0 | 11.2 | 1.73 |
| HSE | 3.959 | 1.039 | 64.5 | 0.019 | 6.97, −2.06/−5.58 | 40.7 | 10.8 | 3.27 |
| B1−WC(PP)[a] | 3.957 | 1.022 | 63.3 | 0.017 | 7.11, −2.08/−5.68 | 33 | 4.9 | 3.22 |
| B1−WC(AE)[a] | 3.962 | 1.015 | 63.2 | 0.015 | 7.08, −2.12/−5.57 | 28 | 4.8 | 3.44 |
| SCAN | 3.985 | 1.029 | 65.1 | 0.017 | 7.11, −2.11/−5.65 | 35.4 | 5.0 | 2.13 |
| Experiments | 3.986[b] | 1.010[b] | 64.0[b] | 0.015[b] | 6.7, −2.4/−4.8[q] | 26[c] | 34(393 K[d]) | 3.27[e] |
| | | | | | | | | 3.38[e] |
| | | | | PbTiO$_3$ | | | | |
| LDA | 3.865 | 1.045 | 60.4 | 0.034 | 7.30, −2.61/−5.99 | 79.8 | 11.6 | 1.47 |
| PBE | 3.844 | 1.240 | 70.4 | 0.058 | 7.37, −2.57/−6.08 | 125.5 | 41.0 | 1.88 |
| HSE | 3.832 | 1.158 | 65.2 | 0.047 | 6.85, −2.55/−5.62 | 114.4 | 38.8 | 3.00 |
| B1−WC(PP)[a] | 3.810 | 1.154 | 63.9 | 0.050 | 6.89, −2.51/−5.76 | 119 | 22.1 | 2.66 |
| B1−WC(AE)[a] | 3.846 | 1.097 | 62.4 | 0.046 | 6.81, −2.51/−5.62 | 103 | 32.4 | 2.83 |
| SCAN | 3.866 | 1.122 | 64.9 | 0.045 | 6.99, −2.63/−5.73 | 105.7 | 24.5 | 2.08 |
| Experiments | 3.880[f] | 1.071[f] | 62.6[f] | 0.040[g] | -- | 57[h] | 67(760 K[d]) | 3.6[k] |
| | | | | | | 75[i] | | |
| | | | | | | 90~100[j] | | |
| | | | | LiNbO$_3$ | | | | |
| LDA | 5.093 | 2.711 | 310.1 | 0.0184 | 9.41, −3.82/−3.51 | 77.9 | 26.8 | 3.33 |
| PBE | 5.203 | 2.712 | 330.9 | 0.0207 | 9.61, −3.97/−3.57 | 84.4 | 30.7 | 3.40 |
| HSE | 5.135 | 2.711 | 317.9 | 0.0207 | 9.13, −3.84/−3.39 | 84.0 | 36.4 | 4.99 |
| SCAN | 5.148 | 2.712 | 320.4 | 0.0201 | 9.29, −3.83/−3.46 | 82.2 | 32.9 | 3.83 |
| Experiments | 5.152[l] | 2.694[l] | 319.0[l] | 0.0193[l] | -- | 71[m] | 128(1483 K[d]) | 3.78[o] |
| | | | | | | 70[n] | | 4.7[p] |

a. Ref. 9
b. Room temperature, Ref. 58
c. Ref. 59
d. Ref. 2, Chapter 1
e. Ref. 60
f. Extrapolated to 0 K from Ref. 61. Also see Ref. 9.
g. Room temperature, Ref. 62
h. Ref. 63
i. Ref. 64; Ref. 65
j. Ref. 66, Chapter 6
k. Ref. 67
l. 293 K, Ref. 68
m. Room temperature, Ref. 69
n. Ref. 70
o. Ref. 71
p. Revised value according to *GW* calculation from Ref. 72
q. Ref. 73

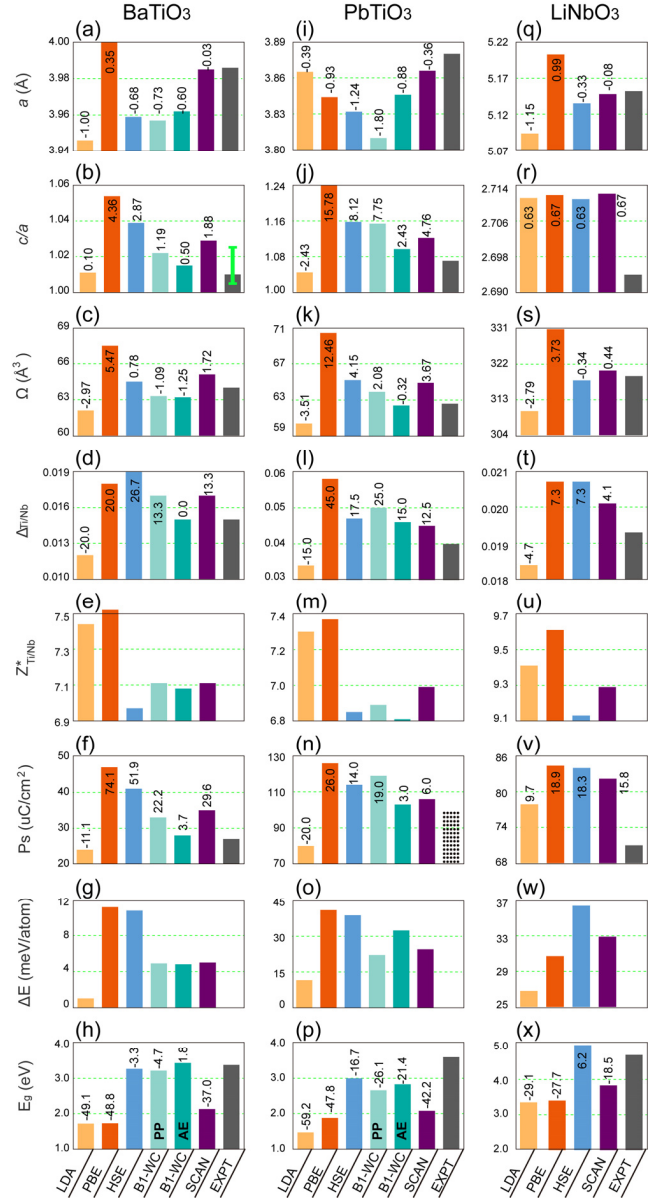

Figure 2. Plots of the properties reported in Table 1 for BaTiO$_3$ (left panel), PbTiO$_3$ (middle panel), and LiNbO$_3$ (right panel). The PP and AE are pseudopotential and all-electron potential, respectively. The annotated values are relative errors in comparison with the experimental values. In subplot (b), an error bar is plotted for the experimental values (see text for details); in subplot (n), the dotted bar indicates that the experimental value of PbTiO$_3$ spontaneous polarization is quite scattered (50 ~ 100 μC/cm$^2$), and the the upper limit of 100 μC/cm$^2$ is used here (see discussions in the main text).

### 3.2.1. Structural properties

We first discuss the structural properties of BaTiO$_3$ [see Figure 2(a-d)] computed from the various XC functionals. First, the lattice parameters (and cell volumes) are underestimated and overestimated by the LDA and PBE, which are consistent with the well-known overbinding and underbinding problems for the two functionals, respectively. The predicted lattice parameters are substantially improved by all the hybrid functionals (i.e., HSE and B1-WC) as well as by the SCAN meta-GGA. Second, the ferroelectric structural distortion, which is closely associated with the O-2$p$ Ti-3$d$ hybridization and measured by the Ti off-center displacement (i.e., $\Delta_{Ti}$) as discussed earlier, is underestimated by LDA

Page **7** of **23**

but overestimated by PBE. It is worth noting that the lattice tetragonality (i.e., $c/a$ ratio) was more often used to measure the strength of structural distortion. The $c/a$ ratio seems to be surprisingly well reproduced by LDA, while largely overestimated by PBE if we take the room temperture experimental value of 1.010[58] as the reference. The observed strong overestimation of *c/a* by PBE, i.e., the supertetragonality problem,[9] is discussed in more details below.

Attention is required when comparing the calculated distortion parameters ($Δ_{Ti}$ and *c/a*) with the experimental measurements. While $Δ_{Ti}$ is difficult to measure accurately, the measured *c/a* ratio has considerable uncertainties, varying from 1.006 to 1.026 [i.e., 1.006,[74] 1.008 at 353K,[75] 1.010 at room temperature,[58] 1.012,[76] 1.018,[77] 1.022,[78] and 1.026[78]]. See error bar in Figure 2(b)]. It is worth noting that the polar distortions can be sensitively influenced by external effects such as experimental temperature and defects. Experimentally, BaTiO$_3$ exists with the tetragonal phase between 278 and 353 K, while at a higher temperature it transforms to the cubic phase. It is thereby reasonable to believe that the reference *c/a* ratio for zero-temperature DFT calculations should be larger than the room-temperature experimental value of 1.010.[58] In this regard, the upper limit of the measured *c/a* value of 1.026[78] (at a lower temperature) should be more appropriate than the room temperature value of 1.010[58] to be used as the reference. Therefore, we argue that the *c/a* ratio is, contrary to the previous belief, notably underestimated by the LDA functional. Consequently, the *super-tetragonality* problem for the PBE functional is not as severe as it was believed to be.

An interesting question can be asked: Why is the tetragonality *c/a* underestimated by LDA but overestimated by PBE? As mentioned earlier, the lattice distortion ($c/a \neq 1$) is driven by the Ti off-center displacement due to the subtle balance between two inequivalent Ti-O bonding interactions along the polarization direction (see Figure 1). Whereas the shorter Ti-O bond has more covalent feature, the longer Ti-O bond has more ionic characteristic. LDA overbinds all chemical bonds and tends to homogenize the electron density, which smooths out the differences between the shorter and longer Ti-O bonds. Therefore, the Ti off-center displacement and thereby the $c/a$ distortion are underestimated. PBE was designed to soften the overestimated bonding calculated by LDA. In general, PBE favors stronger bondings (e.g., covalent and metallic bonds), and thus usually underestimates the relatively weak bonds (the ionic bond in this case). The short Ti-O bond therefore is energetically favored by PBE, leading to the *super-tetragonality*.

We then discuss the results of BaTiO$_3$ calculated from the hybrid functionals and the SCAN meta-GGA. HSE slightly improves the predicted distortions compared with its parent PBE functional. However, the overestimation of the structural distortions is still visible.[26] Substantial improvements in the lattice parameters and structural distortion parameters are achieved by the B1-WC hybrid functional, which was specifically designed for ferroelectric materials.[9,27,28] Interestingly, our SCAN results also agree well with the available experimental results considering their uncertainties and fluctuations as mentioned previously. This is consistent with the fact that SCAN can recognize different chemical bonds.[36]

We finally discuss new features of the structural properties of PbTiO$_3$ and LiNbO$_3$ relative to that of BaTiO$_3$. For PbTiO$_3$ [see Figure 2(i-l)], it is unexpected that the lattice parameter *a* is also underestimated by PBE, although the cell volume is overestimated as found for BaTiO$_3$. It is worth noting that the *c/a* ratio is significantly overestimated (i.e., by 15.78% compared with experiment) by PBE.[79] The calculated structural properties are again largely improved by both B1-WC and SCAN. For LiNbO$_3$ [see Figure 2(q-t)], the general trends of the calculated structural properties are similar to those of BaTiO$_3$. For the *c/a* ratio of LiNbO$_3$, however, all the calculated results have much smaller errors (with overestimations smaller than 1%) in comparison with those of BaTiO$_3$ and PbTiO$_3$. Nevertheless, it should be noticed that the ferroelectricity of LiNbO$_3$ arises from both the Li ordering and Nb displacement, and the connection between the ferroelectricity and the *c/a* distortion is not as straightforward as that of BaTiO$_3$.

### 3.2.2. Ferroelectric properties

The Born effective charge ($Z_s^*$) of atom *s* in periodic solids is defined[2] as $Z_{s,\alpha\beta}^* = \frac{\Omega}{e}\frac{\partial P_\alpha}{\partial u_{s,\beta}}\bigg|_{E=0}$, where $e > 0$ is the charge of an electron and $\Omega$ is the primitive-cell volume. $Z_s^*$ measures the change of polarization ***P*** along the $\alpha$-direction linearly induced by a sublattice displacement $\mathbf{u}_s$ along the $\beta$-direction under zero applied electric field *E*.[2] For ferroelectrics, the dynamical charge $Z_s^*$ can be much greater than the nominal charges for particular ions. Our results [see Table 1 and Figure 2(e,m,u)] are consistent with the fact that the $Z_s^*$ values for the transition metal ions (i.e., Ti in BaTiO$_3$ and PbTiO$_3$, Nb in LiNbO$_3$) are anomalously larger than the corresponding nominal ones due to the *pd* hybridization.[80] The calculated $Z_s^*$ values, for example $Z_{Ti}^*$, of BaTiO$_3$, are substantially overestimated compared with the experimental result [see Table 1]. More interestingly, the values computed by LDA and PBE are larger than that by the hybrid functionals and SCAN. It is well known that the electronic polarizability of ions (for example, a water molecule[26]) is usually overestimated by LDA and PBE because of self-interaction error,[81] which can be partially solved by hybrid functionals. $Z_s^*$ measures the response of an ion's charge to the motion of other ions in the lattice, and is thus closely related to its electronic polarizability. It thus should also likely be overestimated by the LDA and PBE functionals while then corrected by B1-WC. SCAN predicts smaller $Z_s^*$, more reliable than the LDA/PBE results, consistent with the observation we had for the polarizability of water.[26]

For BaTiO$_3$ [see Table 1 and Figure 2(f)], LDA slightly underestimates the spontaneous polarization (with an error −11.1%), much better than PBE (with an error +74.1%). According to the linear approximation, the polarization can be conveniently represented by the polar distortion from structural relaxation times $Z_s^*$ from the electronic structure minimization.[2] Recall that the polar distortion is underestimated and overestimated by a similar amount by LDA and PBE, respectively, while both functionals overestimate $Z_s^*$. As a result, the polarization calculated by LDA is much closer to the experimental value, benefiting from the error cancellation between the underestimation of the polar distortion and the overestimated polarizability as well as $Z_s^*$. By contrast, the polarization is overestimated by PBE due to the accumulation of the overestimations on the two quantities. The best agreement with experiment is achieved by the B1-WC functional with all-electron potential (overestimated by 7.7%); the results from the B1-WC with pseudopotential and the SCAN meta-GGA are slightly worse (with similar overestimations of 20% ~ 30%). When comparing to experiments, we should notice that the experimentally measured polarization could have been reduced by a leakage current,[82] which is usually unavoidable especially in thin films due to defects, grain boundaries, and conduction processes such as Schottky injection or Fowler-Nordheim tunneling.[2] In addition, the polarization decreases as temperature increase since the polar distortion tends to be suppressed when approaching the Curie temperature. Considering these two facts, we argue that the polarization overestimations by SCAN and B1-WC should be reduced after the external effects are extracted from the experimental values, while LDA's underestimation would be enhanced.

We then discuss the calculated polarizations for PbTiO$_3$ and LiNbO$_3$. For PbTiO$_3$ [see Table 1 and Figure 2(n)], a rigorous comparison with the experiment is more difficult since the measured values have significant uncertainties, ranging from 57 to 100 µC/cm$^2$ (e.g., 57,[63] 75[64,65], and 90~100[66] µC/cm$^2$). Nevertheless, a large value should be expected because both the Ti and Pb displacements contribute to the polarization. We point out that the previously reported B1-WC (with all-electron potential) result of 103 µC/cm$^{29}$ and our SCAN result of 105.7 µC/cm$^2$ are almost identical. For LiNbO$_3$ [see Table 1 and Figure 2(v)], all the theoretical results are greater than the experimental values. Again, the experimental data were subject to environmental effects. For example, the polarization of 71 µC/cm$^{2[69]}$ was measured at room temperature, and a sensitive temperature effect was found in Ref. 45.

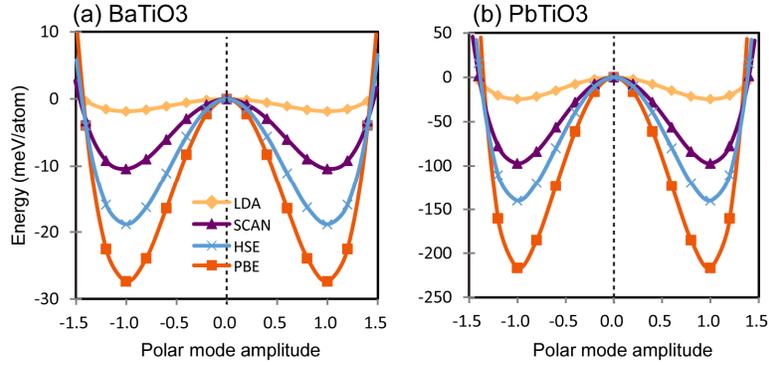

Figure 3. Total energy as a function of the amplitude of the polar distortion for (a) BaTiO$_3$ and (b) PbTiO$_3$ within the LDA, PBE, SCAN, and HSE functionals. In the calculations, the lattice constants are fixed to the relaxed values of the ferroelectric phase, and the internal ionic positions are interpolated along the direction of the polar mode. The double-well depth here is quantitatively slightly different from the energetic difference in Table 1, for which all the structural parameters are fully relaxed.

The energetic stability of the ferroelectric phase with respect to the paraelectric phase is another important property complementary to the structural distortion and electric polarization. Figure 3 [also see Figure 2(g,o)] shows the double-well energy landscape of BaTiO$_3$ and PbTiO$_3$. First of all, it should be noticed that our calculated double-well depth cannot be strictly compared with the energy barrier of the ferroelectric phase transition in reality (i.e., derived from the Curie temperature) because *the ferroelectrics do not switch by a uniform change of the polarization through the paraelectric state,*[83] in which the domain wall plays a critical role. Instead, our aim is to compare the different performances of the adopted XC functionals. As discussed earlier, LDA underestimates the polar distortion. This underestimation artificially reduces the structural difference (and thus energetic difference) between the ferroelectric and the paraelectric phases, which results in too shallow double-well depths. The opposite effect is also true for PBE, which predicts the deepest depths among all the functionals. Finally, SCAN (and HSE) improves over the LDA and PBE functionals, which is qualitatively consistent with the finding on the transition temperature.[30] This is supported by a previous finding that SCAN is accurate for discerning phase stabilities of diversely bonded materials.[26] We mention that the ferroelectric phase of PbTiO$_3$ (with a double-well depth of −220 meV/atom) is more stable than that of BaTiO$_3$ (−28 meV/atom) with respect to their respective paraelectric phases because the polar distortion in PbTiO$_3$ is collectively stabilized by both the Ti-O and Pb-O hybridizations.

### 3.2.3. Lattice dynamics

The paraelectric phase is interesting not only because it is used as a reference in calculating the spontaneous polarization of the ferroelectric phase, but also its lattice dynamics are closely associated with the ferroelectric or antiferrodistortive structural instabilities. In this subsection, we calculate the phonon properties of BaTiO$_3$, PbTiO$_3$, and SrTiO$_3$ within the cubic phase. These materials have been intensively studied in the past,[8,84] and our aim here is to understand the different performances of LDA, PBE, and SCAN. We first evaluate the XC functionals for the prediction of lattice constant, the only independent structural parameter, since it was found that the phonon properties sensitively depend on the cell volume.[85] Table 2 and Figure 4 are results of ten ABO$_3$ (A = Ca, Sr, Ba, Pb, K; B = Ti, Zr, Ta, Nb) systems that are calculated using LDA, PBE, SCAN, as well as the B1-WC functional.[9] The mean error (mean absolute error) for the B1-WC and SCAN results are −0.007 (0.018) and 0.010 (0.013), respectively, compared with −0.033 (0.033) and 0.043 (0.043) for the LDA and PBE data. Obviously, the calculated lattice constants are substantially improved by B1-WC and SCAN.

Table 2. Lattice constants $a$ (Å) of cubic ABO$_3$ (A = Ca, Sr, Ba, Pb, K, K; B = Ti, Zr, Ta, Nb) with space group $Pm\bar{3}m$ that are calculated using the LDA, PBE, B1-WC (with the all-electron potential),[9] and SCAN functionals. Available experimental results are presented. Ratios of the theoretically calculated and experimentally measured lattice constants are also plotted in Figure 4. The mean error (ME) and mean absolute error (MAE) are also given for each theoretical approach.

| Approaches | CaTiO$_3$ | SrTiO$_3$ | BaTiO$_3$ | PbTiO$_3$ | CaZrO$_3$ | SrZrO$_3$ | BaZrO$_3$ | PbZrO$_3$ | KNbO$_3$ | KTaO$_3$ | ME | MAE |
|---|---|---|---|---|---|---|---|---|---|---|---|---|
| LDA | 3.811 | 3.862 | 3.951 | 3.890 | 4.068 | 4.100 | 4.159 | 4.112 | 3.995 | 3.959 | −0.033 | 0.033 |
| PBE | 3.887 | 3.942 | 4.034 | 3.970 | 4.138 | 4.175 | 4.236 | 4.187 | 4.062 | 4.028 | 0.043 | 0.043 |
| B1-WC[a] | 3.834 | 3.880 | 3.971 | 3.901 | 4.111 | 4.138 | 4.195 | 4.148 | -- | 3.971 | −0.007 | 0.018 |
| SCAN | 3.852 | 3.909 | 4.003 | 3.935 | 4.108 | 4.144 | 4.208 | 4.155 | 4.033 | 3.982 | 0.010 | 0.013 |
| Experiment | 3.836[b] | 3.905[c] 3.89[d] | 4.001[e] | 3.93[f] | 4.12[h] | 4.109[i] | 4.192[j] | 4.13[k] | 4.02[g] | 3.988[l] | -- | -- |

a. Ref. 9
b. 600 K, Ref. 86
c. Room temperature, Ref. 87
d. Extrapolated to 0 K from Ref. 87,88
e. Ref. 87
f. Ref. 61
g. Ref. 89
h. Ref. 90
i. Ref. 91
j. Ref. 92
k. Extrapolated to 0 K from Ref. 93
l. Ref. 94

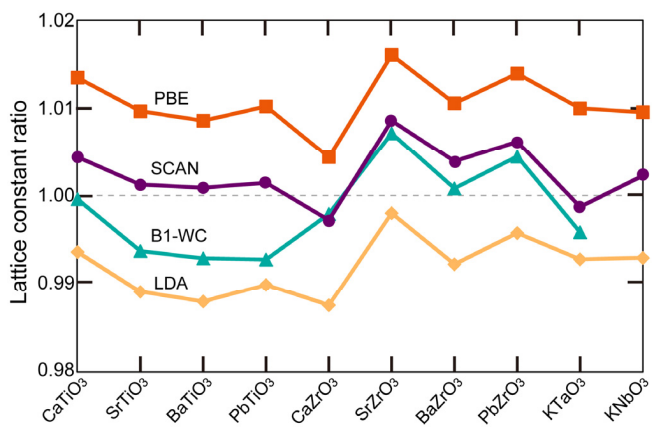

Figure 4. Ratios of the theoretically calculated and experimentally measured lattice constants from Table 2

Table 3 shows the Γ point vibrational frequencies (of transverse optical modes) calculated by LDA, PBE, and SCAN, and the previously reported hybrid functionals result as well. Figure 5 also shows the phonon dispersions spanning the Brillouin zone. We first discuss the results for BaTiO$_3$ calculated with crystal structures relaxed by the underlying functionals [see Table 3 and Figure 5(a)]. In general, the phonon frequencies computed by LDA agree reasonably well with the experimental values; however, PBE tends to predict softer phonon modes (i.e., with lower frequencies) than the experiments. The error cancellation effect for LDA in the polarization calculation (see Section 3.2.2) encourages us to make a similar analysis for the phonon property. On the one hand, the well-known overbinding tendency by LDA can lead to an overestimation of bonding strength,[95] which tends to over-stiffen the phonon modes. On the other hand, the overestimation of the electronic polarizability by LDA yields an artificially enhanced resonant feature[84] of the bonding interactions, which helps to soften the phonon modes.[96] Due to the above error cancellation, LDA predicts relatively accurate phonon frequencies. For the PBE calculations, on the contrary, both the underestimation of the bond strength and the overestimation of the atomic polarizability artificially reduce the phonon frequency. Again, the results calculated by both B1-WC and SCAN agree much better with the experiments. Finally, the imaginary frequencies [of the $\Gamma_{15}$(TO1) modes] computed by LDA are higher than the B1-WC and the SCAN results. Since the soft mode measures the structural instability, the results here suggest that LDA tends to overstabilize the cubic phase, which is consistent with the fact that LDA underestimates the polar distortion (see Section 3.2.2).

Table 3. The $\Gamma$ point phonon frequencies (cm$^{-1}$) of cubic BaTiO$_3$, PbTiO$_3$, and SrTiO$_3$ calculated by the LDA, PBE, SCAN, and hybrid functionals. The three $\Gamma_{15}$ modes are triply degenerate, and the $\Gamma_{25}$ mode is triply degenerate. Only the transverse optical (i.e., TO) modes are presented in this table in order to compare with the available previous reports.[9,85] For the hybrid functionals, B1-WC is used for BaTiO$_3$ and PbTiO$_3$; HSE is used for SrTiO$_3$. For the B1-WC functional, both the all-electron and pseudopotential (values in parenthesis) results are given. The lattice parameters are relaxed using each functional.

| Modes | LDA | PBE | B1-WC[a]/HSE[d] | SCAN | Experiment |
|---|---|---|---|---|---|
| BaTiO$_3$ | | | | | |
| $\Gamma_{15}$(TO1) | −145 | −247 | −145(−213)[a] | −220 | -- |
| $\Gamma_{15}$(TO2) | 186 | 169 | 195( 195)[a] | 183 | 182[b] |
| $\Gamma_{25}$ | 290 | 285 | 299( 298)[a] | 290 | 306[b,c] |
| $\Gamma_{15}$(TO3) | 479 | 452 | 482( 476)[a] | 476 | 482[b] |
| PbTiO$_3$ | | | | | |
| $\Gamma_{15}$(TO1) | −135 | −182 | −146(−196)[a] | −171 | -- |
| $\Gamma_{15}$(TO2) | 127 | 87 | 138( 120)[a] | 109 | -- |
| $\Gamma_{25}$ | 224 | 227 | 231( 229)[a] | 227 | -- |
| $\Gamma_{15}$(TO3) | 509 | 463 | 513( 506)[a] | 498 | -- |
| SrTiO$_3$ | | | | | |
| $\Gamma_{15}$(TO1) | 52 | −130 | −74[d] | 80 | 91.7,[e] 88±1[f] |
| $\Gamma_{15}$(TO2) | 172 | 146 | 162[d] | 173 | 169±3,[e] 175±2[f] |
| $\Gamma_{25}$ | 225 | 232 | 250[d] | 250 | 265±5,[e] 266±3[f] |
| $\Gamma_{15}$(TO3) | 560 | 510 | 533[d] | 546 | 547±3,[e] 545±1[f] |

a. B1-WC functional, Ref. 9
b. Ref. 97
c. This value was measured in tetragonal phase from Ref. 97
d. HSE functional, Ref. 85
e. Neutron scattering, measured at 297 K, Ref. 98
f. Hyper-Raman scattering, measured at room temperature, Ref. 99

The general discussions on BaTiO$_3$ can also be applied to PbTiO$_3$ [see Table 3 and Figure 5(b)]. In addition to the zone-center $\Gamma$–instability leading to polar distortion (i.e., a polar mode), PbTiO$_3$ (like SrTiO$_3$) also has a zone-boundary $R$-instability associated with TiO$_6$ octahedral rotations (i.e., an antiferrodistortive mode).[84]. In particular, the predicted $R$-instability is more unstable within LDA than that within PBE, which is contrary to their performances for $\Gamma$-instability. The opposite ordering of the $R$ and $\Gamma$ instabilities with respect to LDA and PBE is related to the calculated cell volumes: A greater volume tends to destabilize (soften) the polar mode while stiffening the rotational mode.[100,101] To be specific, PBE overestimates the cell volume, and it thereby strongly favors the $\Gamma$-instability; LDA underestimates the cell volume, and it more favors the $R$-instability instead. Considering the strong competition between the polar and antiferrodistortive modes,[102] it is important to use accurate lattice parameters for the phonon property calculations. In this regard, we argue that SCAN should significantly improve the predicted phonon dispersions since it can more reliably reproduce both the lattice parameters (see Table 2 and Figure 4) and the electronic polarizability [see Section 3.2.2].

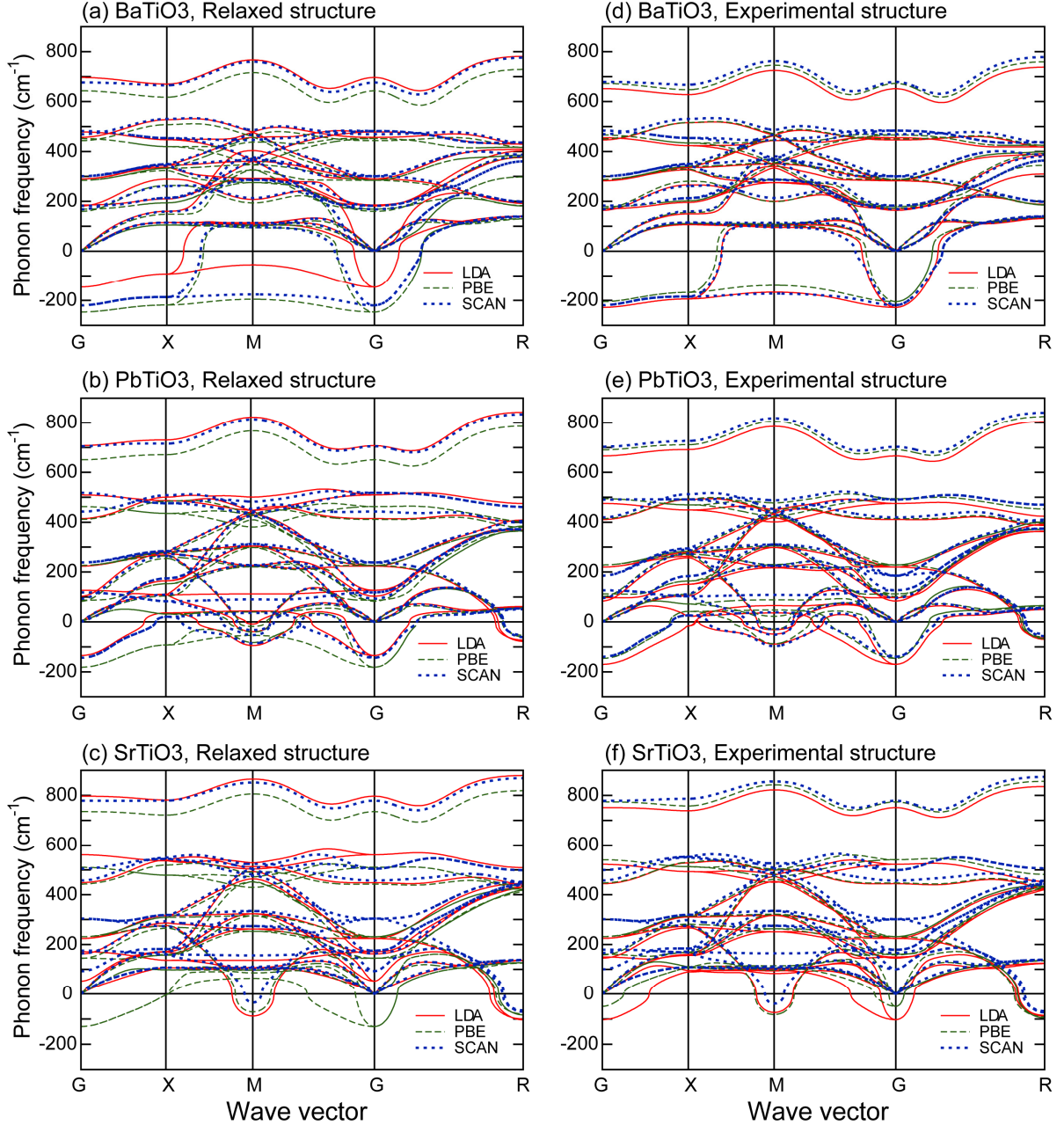

**Figure 5.** Phonon dispersion relations of cubic BaTiO$_3$, PbTiO$_3$, and SrTiO$_3$ calculated by LDA, PBE, and SCAN. In the left panel the crystal structures are fully relaxed by each functional (see Table 2), and in the right panel the experimental lattice parameters of $a$(BaTiO$_3$) = 4.001 Å,[87] $a$(PbTiO$_3$) = 3.93 Å,[61] and $a$(SrTiO$_3$) = 3.905 Å[87] are used. The splitting between longitudinal optical (LO) and transverse optical (TO) phonons (i.e., the LO-TO splitting) is considered in the calculation.

The reliable calculation of SrTiO$_3$ phonon properties is more challenging. The cubic phase has both the ferroelectric and antiferrodistortive instabilities according to the previous first-principles calculations,[103,104] contrary to the fact that the cubic phase is stable from experimental observations.[98,99] Our results from the three XC functionals are very different [see Table 3 and Figure 5(c)]. First, the Γ$_{15}$(TO1) mode is stable within both the LDA and the SCAN functionals, while it is strongly unstable within the PBE calculation. Second, the calculated M and R instabilities by SCAN are not as strong as those by LDA and PBE. In summary, although the Γ mode frequencies are well reproduced by SCAN, the predicted M and R instabilities still disagree with the experiments. We point out that the discrepancies might be caused by some external effects, such as structural disorder and finite temperature, which are not considered here. For instance, it was recently found that the observed cubic phase is actually dynamic averages of the tetragonal structures.[105] It was also found that the imaginary modes are caused by finite-temperature effects[106] According to their findings, the imaginary frequencies can be precluded by using force constants either calculated from the tetragonal phases,[105] or with incorporated temperature effect.[106]

Conventionally, LDA is usually used for the phonon property calculations with, however, the experimental lattice constants to avoid the lattice underestimation problem.[8] To evaluate this procedure, we also calculate phonon dispersions by LDA, PBE, and SCAN but with the experimental lattices [see Figure 5(d-f)]. The most obvious change is that the polar instability predicted by LDA is strongly affected by choice of the lattice constants because the LDA lattices have the largest deviations from the experimental ones (see Table 2 and Figure 4). In SrTiO$_3$, moreover, the Γ instability predicted by LDA is so unstable that it is incompatible with the fact that the ferroelectric instability can be suppressed by weak zero-point quantum fluctuations.[107] Since the experimental lattice constants are well reproduced by SCAN, the choice of lattice constants has negligible effects on the SCAN phonon frequencies. We conclude that SCAN gives the most consistent descriptions on the lattice dynamic properties of these materials.

### 3.3. Hydrogen-bonded ferroelectrics: inorganic KH$_2$PO$_4$ and organic PhMDA

#### 3.3.1. Structural properties

In this subsection, we investigate the structural properties of the inorganic KH$_2$PO$_4$ and the organic PhMDA (see Table 4), of which the ferroelectric properties are mainly determined by hydrogen bonding interactions. First of all, we stress that the lattice parameters and cell volumes are best predicted by SCAN due to its ability to capture the intermediate-range van der Waals (vdW) interaction, which is important for these materials.[26] Nevertheless, the PhMDA cell volume calculated by SCAN still has a considerable discrepancy (with an error -4.4%) relative to the experimental value, which is larger than that for KH$_2$PO$_4$ (with an error +0.4%). A possible reason is that the finite-temperature effect (measured at 293 K[108]) may not be neglected, considering the light masses of constituent elements. In the following paragraphs, we focus on an internal structural parameter, i.e., the difference between a relatively strong proton-donor bond (denoted by H–O) and a relatively weak proton-acceptor (denoted by H⋯O) in an O−H⋯O structure (see Figure 1). We will show that the difference plays a key role in determining the ferroelectric structural distortion.

The proton location in an O−H⋯O structure is determined by the delicate competition between two attractive interactions, i.e., the strong H-O covalent bonding and the relatively weak H⋯O hydrogen bonding. We define a parameter $\eta \equiv R_{\text{H}\cdots\text{O}}/R_{\text{H}-\text{O}}$ to measure the difference of the two bond lengths, which is closely related to the ferroelectric structural distortion. For both KH$_2$PO$_4$ and PhMDA, LDA tends to put the proton at the center of the O−H⋯O structure (i.e., $\eta \to 1$), indicating that LDA tries to make the least inhomogeneous electron density around the central proton. The underestimated distortion is similar to the previous finding in BaTiO$_3$ and PbTiO$_3$ (see Section 3.2.1). In fact, the proton in KH$_2$PO$_4$ is exactly placed at the center (i.e., $\eta = 1$), rendering two equivalent hydrogen-oxygen bonds known as *symmetric hydrogen bonds*.[109] These bonds, which have some covalent character, are much stronger than the 'normal' H⋯O hydrogen bonds.[52,110] They are usually found in materials at high-pressure such as in ice X[111,112] and potassium hydrogen maleate.[113-115] The hydrogen bond overbinding by LDA was also found in water.[26]

Table 4. Structural, ferroelectric, and electronic properties of KH$_2$PO$_4$ and PhMDA with the ferroelectric phase. Note that the experimental structural parameters were measured at 115 K and 293 K for KH$_2$PO$_4$ and PhMDA, respectively.

| Properties | KH$_2$PO$_4$ | | | | PhMDA | | | |
| --- | --- | --- | --- | --- | --- | --- | --- | --- |
| | LDA | PBE | SCAN | Experiments | LDA | PBE | SCAN | Experiments |
| Space group | $I\bar{4}2d$ | $Fdd2$ | $Fdd2$ | $Fdd2$[a] | $Pna2_1$ | $Pna2_1$ | $Pna2_1$ | $Pna2_1$[f] |
| Lattice $a$ (Å) | 10.32 | 10.80 | 10.53 | 10.53[a] | 7.06 | 8.11 | 7.34 | 7.683[f] |
| Lattice $b$ (Å) | 10.32 | 10.71 | 10.50 | 10.47[a] | 16.73 | 18.28 | 17.31 | 17.16[f] |
| Lattice $c$ (Å) | 6.78 | 7.11 | 6.93 | 6.93[a] | 5.43 | 5.38 | 5.51 | 5.553[f] |
| Volume $\Omega$ (Å$^3$) | 721.5 | 822.2 | 766.6 | 763.6[a] | 641.3 | 798.3 | 700.2 | 732.1[f] |
| Bond length, $R_{OO}$ (Å) | 2.403 | 2.508 | 2.492 | 2.491[a] | 2.400 | 2.552 | 2.494 | 2.604[f] |
| Bond length, $R_{O-H}$ (Å) | 1.201 | 1.057 | 1.041 | 1.059[b] | 1.197 | 1.049 | 1.041 | -- |
| Bond length, $R_{O\cdots H}$ (Å) | 1.201 | 1.451 | 1.451 | 1.441[b] | 1.204 | 1.505 | 1.453 | -- |
| Bond length ratio, $R_{O\cdots H}/R_{O-H}$ | 1.000 | 1.372 | 1.394 | 1.360[b] | 1.006 | 1.435 | 1.396 | -- |
| Bond angle, $\angle$O−H$\cdots$O (deg) | 179.5 | 178.1 | 179.9 | 172.2[b] | 179.6 | 176.9 | 178.5 | -- |
| Polarization, $P_s$ (μC/cm$^2$) | 0.0 | 5.49 | 5.91 | 5.12[c] | 0.13 | 6.58 | 6.94 | 9[f] |
| Stability, $E_{PF}-E_{FE}$ (meV/atom) | 0.0 | 17.0 | 11.6 | 11(122 K[d]) | 0.1 | 2.6 | 2.2 | 31(363 K[f]) |
| Band gap, $E_g$ (eV) | 5.82 | 5.39 | 6.22 | 5.90[e] | 1.82 | 2.47 | 2.42 | -- |

a. 115 K, Ref. 116
b. Room temperature, Ref. 117
c. 93 K, Ref. 118
d. Ref. 119 and Ref. 120
e. Ref. 121. Note, the band gap is expected to be strongly influenced by the exciton effect.
f. 293 K, Ref. 108

Interestingly, the structural distortion parameter $\eta$ is similar from PBE and SCAN, and they agree well with the experimental results. The good performance of the PBE functional for the hydrogen bond strength and thereby the parameter $\eta$ has been well recognized.[26,122] There are two error sources from PBE for the O−H$\cdots$O structure. The self-interaction error of PBE over-delocalizes the lone-pair electrons of O$^{2-}$ ions and thus overstabilizes the H$\cdots$O bond, while PBE misses the intermediate-range vdW interaction between two adjacent oxygen ions in two nearby O−H$\cdots$O structures. These two errors cancel each other largely, resulting in good performance of PBE for hydrogen-bonded structures. SCAN improves significantly over PBE for the intermediate-range vdW interactions and considerably for the self-interaction error.[26] It, however, still experiences some self-interaction errors. SCAN thereby yields a similar level of accuracy for the structural properties and the ferroelectric polarization as the PBE functional, while SCAN gives better bond lengths and cell volumes. Nevertheless, the hydrogen bond strength is overestimated by both SCAN and PBE due to the self-interaction error.[26] It is still challenging to accurately calculate the structural properties using the tested XC functionals due to the delicate competition between the strong H-O covalent bonding and the weak H$\cdots$O hydrogen bonding. Finally, it is worth noting that the nuclear quantum effect should also be considered for these materials since Hydrogen is a very light atom, which further weakens the strength of the hydrogen bond in water.

### 3.3.2. Ferroelectric properties

Table 4 shows the spontaneous polarizations for the two hydrogen-bonded materials. In KH$_2$PO$_4$, LDA predicts a vanishing polarization. This is consistent with the incorrectly predicted hydrogen-bonded structure in which LDA tends to place the proton in the middle of two oxygen atoms resulting in a weakly broken inversion symmetry. The calculated polarizations by PBE ($P_s$ = 5.49 μC/cm$^2$) and SCAN (5.91 μC/cm$^2$) agree well with the experimental value (5.12 μC/cm$^2$[118]).

For PhMDA, the polarization is again strongly underestimated by LDA (0.13 μC/cm$^2$). The PBE and SCAN significantly improve the prediction by yielding polarizations of 6.58 μC/cm$^2$ and 6.94 μC/cm$^2$, respectively, in comparison with the experimental value of 9 μC/cm$^2$.[108] It is also interesting to briefly discuss the structural origin of the spontaneous polarization of KH$_2$PO$_4$ and PhMDA. The polarization of KH$_2$PO$_4$, which is along the out-of-plane direction [see Figure 1(c)], is mainly (i.e., amounting to 66% within the PBE calculation) induced by the in-plane proton displacement via not only the nearest-neighbor hydrogen-oxygen interactions but also the further-neighbor interactions.[123] Indeed, the calculated Born effective charge tensor has anomalously large values for both the diagonal and off-diagonal components.[56] The remaining part of the polarization is from the accompanied phosphorus displacement. By contrast, the O–H···O structures in PhMDA have nonzero components along the polarization direction [see Figure 1(d)]. Therefore, the hydrogen bond structures make a more significant contribution (amounting to 80%) to the polarization of PhMDA.[23]

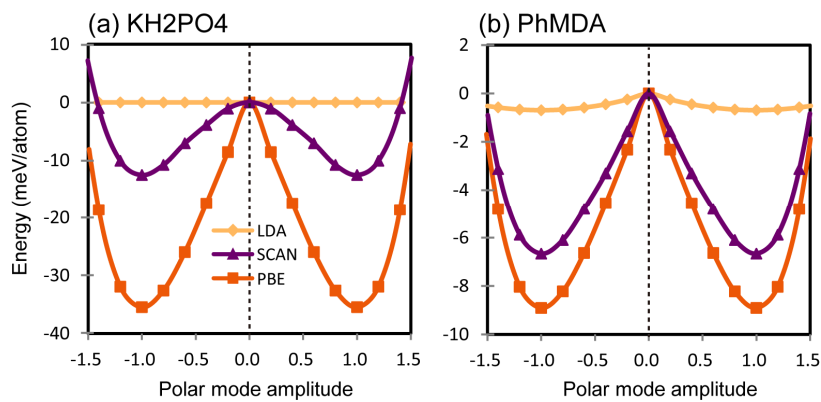

Figure 6. Total energy as a function of the amplitude of the polar distortion between centric and polar configurations for (a) KH$_2$PO$_4$ and (b) PhMDA within the LDA, PBE, and SCAN functionals. In the calculations, the lattice constants are fixed to the relaxed values of the low-temperature structure, and the internal ion positions are interpolated. The double-well depth here is quantitatively slightly different from the energetic difference in Table 4, for which all the structural parameters are fully relaxed.

The energetics of KH$_2$PO$_4$ and PhMDA predicted by LDA, PBE, and SCAN (see Figure 6) in general follow the same trend as for BaTiO$_3$ and PbTiO$_3$ (see Figure 3). Nevertheless, it should be noticed that the double-well depths for the hydrogen-bonded systems are much shallower within the LDA functional. For KH$_2$PO$_4$, the double-well feature is even absent by the LDA calculation, indicating that the ferroelectric phase (space group $Fdd2$) cannot be stabilized; instead, the paraelectric phase (space group $I\bar{4}2d$) is incorrectly predicted to be more stable (see Table 4) by LDA.

### 3.4. Multiferroics with open non-$d^0$-shell ions: proper BiFeO$_3$ and improper YMnO$_3$

In this section, we evaluate the XC functionals for the structural and ferroelectric properties of two prototypical multiferroic materials, i.e., the proper ferroelectric BiFeO$_3$ and the improper hexagonal YMnO$_3$. There is an additional difficulty for the theoretical investigations because of the transition metal ions with open 3$d$ shells. The band gaps might fail to be opened by the conventional LDA and PBE functionals, which prohibits the polarization calculations using the modern theory of polarization.[7] In fact, some other fundamental properties (e.g., band structure, magnetic moment, and lattice dynamics) may not be well described either. To better deal with the 3$d$ electrons, an empirical on-site Coulomb $U$ can be combined with LDA, PBE, and SCAN, resulting in the DFT+$U$ approach.[124] Alternatively, the hybrid functionals were also used for these materials. Improved descriptions of the structural and ferroelectric properties are expected for those approaches.

Table 5. Structural properties (lattice $a$, $c/a$ ratio, and unit cell volume $\Omega$), ferroelectric properties (spontaneous polarization, and the energetic difference between the ferroelectric and paraelectric phases), and electronic properties (band gap) of $R3c$ BiFeO$_3$ and $P6_3cm$ YMnO$_3$. The hexagonal cell of BiFeO$_3$ is used for the calculation. Some results are also plotted in Figure 7 for easier comparison. For YMnO$_3$, the polarization calculation by some approaches is unavailable because insulating band structures are required by the modern theory of polarization. Some results for BiFeO$_3$ are reproduced from Ref. 26.

| Approaches | Lattice $a$ (Å) | Lattice ratio $c/a$ | Volume $\Omega$ (Å$^3$) | Polarization $P_s$ (μC/m$^2$) | $E_{PF}-E_{FE}$ $\Delta E$ (meV/atom) | Band gap $E_g$ (eV) |
|---|---|---|---|---|---|---|
| **BiFeO$_3$** | | | | | | |
| LDA | 5.478 | 2.425 | 345.1 | 98.9 | 61.7 | 0.34 |
| PBE | 5.618 | 2.493 | 382.7 | 104.8 | 120.4 | 1.05 |
| SCAN | 5.562 | 2.482 | 369.8 | 102.7 | 63.3 | 1.89 |
| HSE[a] | 5.576 | 2.499 | 375.1 | 110.3 | -- | 3.4 |
| B1-WC[b] | 5.556 | 2.485 | 369.0 | -- | -- | 3.0 |
| LDA+$U$ (2 eV) | 5.497 | 2.453 | 352.9 | 94.8 | 62.3 | 1.40 |
| PBE+$U$ (2 eV) | 5.623 | 2.500 | 384.8 | 100.3 | 58.5 | 1.76 |
| SCAN+$U$ (2 eV) | 5.565 | 2.485 | 371.1 | 99.3 | 63.3 | 2.46 |
| Experiments | 5.579[c,d] | 2.486[c,d] | 373.8[c,d] | 100[e] | 95(1100K[f]) | 2.74[g] |
| **YMnO$_3$** | | | | | | |
| LDA | 6.019 | 1.877 | 354.4 | -- | 23.6 | 0.00 |
| PBE | 6.185 | 1.856 | 380.3 | -- | 19.6 | 0.00 |
| SCAN | 6.133 | 1.860 | 371.6 | -- | 19.7 | 0.00 |
| HSE | 6.147 | 1.849 | 371.9 | 7.2 | 20.7 | 1.27 |
| B1-WC[h] | 6.144 | 1.843 | 370.2 | -- | -- | 1.0 |
| LDA+$U$ (7.5 eV) | 6.088 | 1.870 | 365.4 | 7.4 | 23.8 | 0.25 |
| PBE+$U$ (7.5 eV) | 6.243 | 1.865 | 393.0 | -- | 20.7 | 0.00 |
| SCAN+$U$ (2 eV) | 6.149 | 1.858 | 374.2 | 6.0 | 20.5 | 0.23 |
| Experiments | 6.121[i] | 1.864[i] | 370.1[i] | 5.5[j] | 109(1258K[k]) | 1.28[l], 1.55[m] |

a. Ref. 125. Note, the original reference data are for a rhombohedral cell, and they are transformed to the hexagonal cell here.
b. Ref. 27. Note, the original reference data are for a rhombohedral cell, and they are transformed to the hexagonal cell here.
c. 294 K, Ref. 126
d. Room temperature, Ref. 127
e. Ref. 128
f. Ref. 129
g. Ref. 130
h. Ref. 28
i. 10 K, Ref. 131
j. Ref. 132
k. Ref. 133
l. Room temperature, unknown phase, Ref. 134
m. 4 K, unknown phase, Ref. 134

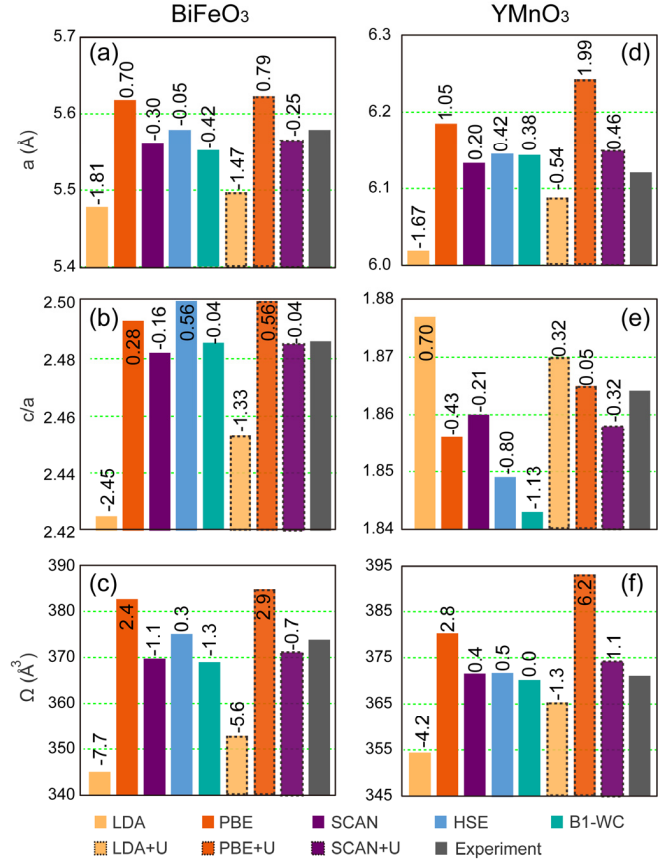

**Figure 7.** Plots of the lattice parameter *a*, lattice distortion *c/a*, and unit cell volume *Ω* reported in Table 5. The annotated values are the errors of the calculated results relative to the experimental values.

We discuss the structural properties of magnetic $BiFeO_3$ [i.e., the lattice constant, $c/a$ ratio, and cell volume; see Table 5 and Figure 7(a-c)] and compare them with those of the non-magnetic $LiNbO_3$, which has the same $R3c$ symmetry. The general trend that the LDA (PBE) underestimates (overestimates) the lattice parameters of $LiNbO_3$ is also observed for $BiFeO_3$. Nevertheless, the errors are more significant for $BiFeO_3$. For example, relative errors of $BiFeO_3$ cell volume are −7.7% (+2.4%) from the LDA (PBE) calculation, in comparison with −2.8% (+3.7%) for $LiNbO_3$. This feature implies that the Fe-$3d$ states play an important role for the structural properties of $BiFeO_3$. The combination of a Coulomb *U* (*U* = 2 eV)[135] with the LDA functional, which describes better the Fe-$3d$ correlation effect, effectively reduces the volume underestimation error by LDA. Similarly, the hybrid functionals also well reproduce the structural parameters because the correlation effect is better calculated by the exact exchange. Finally, it is surprising that SCAN results agree very well with the B1-WC calculations and the experimental values, which might suggest that the Fe-$3d$ states are reasonably described by SCAN as well.

The spontaneous polarizations of $BiFeO_3$ calculated by LDA, PBE, and SCAN follow similar trends for $LiNbO_3$. An interesting feature for the $BiFeO_3$ calculations, however, is that the polarization becomes smaller when combining with the Coulomb *U* (see Table 5). For example, it decreases by an amount of 4.1 µC/cm² from the $P_s^{LDA} = 98.9$ to the $P_s^{LDA+U} = 94.8$ µC/cm². The polarization reduction is caused by the differences found for the structural and electronic properties within the two approaches. Since dipole moment can be theoretically decomposed into the ionic contribution (related to the crystal structure) and the electronic contribution (numerically measured by the Born effective charge), the electronic part can be disentangled if we use the same crystal structure in the calculations. For example, the *U* effect on LDA is 2.29 μC/cm$^2$ (i.e., $P_s^{\text{LDA}} = 105.57$ and $P_s^{\text{LDA+U}} = 103.28$ μC/cm$^2$ ) at the experimental structural parameters. Theoretical description of the Fe-3$d$ states is important for the ferroelectric properties because the open shells influence not only the structural distortion but also the electronic rehybridization in generating the spontaneous polarization.

Finally, we discuss the structural and ferroelectric properties of YMnO$_3$ [see Table 5 and Figure 7(d-f)], which were usually calculated by the DFT+$U$ approaches[136,137] and the hybrid functionals[28] in order to open the band gap. For the same purpose, we use a $U$ = 7.5 eV for the LDA+$U$ and PBE+$U$ methods, but a smaller value of 2 eV for the SCAN+$U$. According to our calculations, whereas the lattice constant and cell volume are strongly underestimated by LDA, these parameters are notably improved within the LDA+$U$ approach. A possible reason is that the band gap is opened by the latter approach. By contrast, the structural parameters are much more overestimated by the PBE+$U$ approach. For the hybrid functionals, the $c/a$ ratio is strongly underestimated. Overall, the structural properties are best reproduced by the SCAN functional and the SCAN+$U$ approach. For the polarization, the calculation is only feasible for a few approaches (i.e., HSE, LDA+$U$, and SCAN+$U$), and the best agreement with experiment is achieved by the SCAN+$U$ ($U$ = 2 eV) approach.[79]

## 4. CONCLUSION

In this paper, we evaluated the performance of LDA, PBE, the hybrid functionals HSE and B1-WC, and the recently developed SCAN meta-GGA for structural and electric properties of several prototypical ferroelectric (BaTiO$_3$, PbTiO$_3$, and LiNbO$_3$; KH$_2$PO$_4$ and PhMDA) and multiferroic materials (BiFeO$_3$ and YMnO$_3$) with diverse bonding interactions (e.g., the covalent, ionic, and hydrogen bondings). LDA works well for the spontaneous polarization of the inorganic systems, which benefits from the error cancellation between the underestimation of structural distortion but the overestimation of the Born effective charge (related to the electronic polarizability). LDA also works well for the lattice dynamics (i.e., the phonon frequencies) due to counteracting effects of phonon over-stiffening by overestimating the bond strength and the phonon over-softening by overestimating the bonding resonant characteristic (again, related to the electronic polarizability). These error cancellations have made the LDA a preferred functional in studying the ferroelectric materials for a long time. On the contrary, PBE strongly overestimates the polarization of inorganic systems because of the overestimations of both the structural distortion and the electronic polarizability. Meanwhile, it usually gives too low phonon frequencies because it underestimates the bond strength and overestimates the bonding resonant characteristic. For hydrogen-bonded systems, LDA severely underestimates the polar distortion because of the overbinding problem, and PBE seems to be more reliable.

The system-dependent performances of LDA and PBE are related to the diverse bonding interactions in the ferroelectric materials. Ferroelectric distortions are driven by local asymmetric structural distortion related to two dissimilar bonding interactions, and their delicate competition determines the distortion magnitude. Because of the ability to recognize the various bonds, the SCAN meta-GGA is a universally effective approach for all the selected ferroelectric materials. First, it significantly improves the calculated structural properties, including both the lattice constants and the structural distortions. Second, it works better than LDA/PBE for the electronic polarizability, which is important for calculating the ferroelectric properties. Finally, the fundamental electronic properties of multiferroic materials are also better described by SCAN as well as by its combination with a small Coulomb $U$. The SCAN meta-GGA is as accurate as or even more accurate in some cases than the B1-WC hybrid functional, which was specifically designed for the ferroelectric materials.

# ACKNOWLEDGEMENTS

This paper was mainly supported by the Air Force Office of Scientific Research under FA9550-13-1-0124 (Y.Z. and X.W. designed the project and wrote the paper; Y.Z. carried out the DFT calculations and analysis.). It was partially supported as part of the Center for the Computational Design of Functional Layered Materials, an Energy Frontier Research Center funded by the U.S. Department of Energy (DOE), Office of Science, Basic Energy Sciences under Award No. DE-SC0012575 (J.S. and J.P.P. invented the SCAN functional and revised the manuscript) and by the National Science Foundation (NSF), Division of Materials Research under Award No. DMR-1552287 (Y.Z. and X.W. carried out analysis of the hydrogen bond ferroelectric materials). This research also used resources of the National Energy Research Scientific Computing Center (NERSC), a DOE Office of Science User Facility supported by the Office of Science of the U.S. Department of Energy under Contract No. DE-AC02-05CH11231. This research was supported in part by the NSF through major research instrumentation Grant No. CNS-09-58854.